\documentclass[cernpreprint, atlasdraft=false, USenglish, texmf, orcidlogo]{atlasdoc}
\usepackage[subcaption]{atlaspackage}
\usepackage[backref=false]{atlasbiblatex}

\usepackage{atlasphysics}
\graphicspath{{logos/}{figures/}}

\AtlasTitle{Measurement of charged-particle production in $\sqrt{s_\text{NN}}=9.62$ TeV proton-oxygen collisions as a probe of cosmic-ray air showers with the ATLAS detector}

\AtlasVersion{3.1}

\AtlasAbstract{%
This Letter presents a measurement of prompt charged-particle production in proton-oxygen interactions at $\sqrt{s_\text{NN}}=9.62$ TeV center-of-mass energy with the ATLAS detector, corresponding to 634~$\mu$b$^{-1}$ of integrated luminosity.
A total of 246 million selected events have at least one track with transverse momentum $p_\text{T}> 500$ MeV and pseudorapidity $|\eta|<2.5$.
The measured fiducial proton-oxygen cross section is $\sigma_\text{fid.}^{p\text{O}}=396 \pm 6~(\text{exp.}) \pm 9~(\text{lumi.})~\text{mb}$ and the extrapolated inelastic proton-air cross section is $\sigma^{p+\text{air}}_\text{inel.} = 406 \pm 6~(\text{exp.}) \pm 9~(\text{lumi.}) \pm 28~(\text{th.})~\text{mb}$.
Measurements of charged-particle multiplicity, $p_\text{T}$, and $\eta$ distributions are an order-of-magnitude more precise than differences between hadronic-interaction models.
These results enable improved modeling of cosmic-ray air showers, which is important for astroparticle physics.
}

\AtlasRefCode{STDM-2025-08}

\PreprintIdNumber{CERN-EP-2026-106}

\AtlasJournalRef{Phys. Rev. Lett. 137 (2026) 121901}
\AtlasDOI{10.1103/f3nk-5lt9}

\AtlasCoverEgroupAnalysisTeam{atlas-stdm-2025-08-analysis-team@cern.ch}

\hypersetup{pdftitle={ATLAS document},pdfauthor={The ATLAS Collaboration}}

\begin{document}

\maketitle

Revealing the origins of cosmic rays is a central goal in astroparticle physics~\cite{Bhattacharjee:1999mup,blumer2009,AlvesBatista:2019tlv,Meszaros:2019xej,Coleman:2022abf}.
The steeply falling energy-dependent flux precludes direct spaceborne measurements above PeV energies, requiring indirect atmospheric observations via ground-based detectors~\cite{HiRes:2007lra,Apel:2013uni,Aab:2014pza,TelescopeArray:2023sbd,Albert:2024aaa}.
Studying PeV cosmic rays is key to elucidating their mass composition~\cite{HiRes:1999ioa,PierreAuger:2016qzj,Dembinski:2025nmp}, the transition from galactic to extragalactic sources~\cite{Swordy:2002df,Hoerandel:2002yg,DeDonato:2008wq,TelescopeArray:2014tsd,Aab:2017tyv}, and the origins of planetary habitability~\cite{Ferrari2009,ATRI2014186,Nojiri:2024lnc,Kamenetskiy2025}.
Collider detectors also measure atmospheric-muon fluxes from cosmic-ray showers~\cite{Avati:2000mn,Achard:2004ws,Abdallah:2007fk,CMS-MUO-10-001,ALICE:2015wfa,ALICE:2024yqj}.
Reconstructing cosmic rays from air showers demands accurate simulation of particle interactions in Earth's atmosphere at \TeV center-of-mass energies~\cite{Alvarez-Muniz:2002lml,Drescher:2002cr,Matthews:2005sd,Engel:2018akg}.
Quantum chromodynamics governs the initial high-energy interaction, whose nonperturbative processes are difficult to calculate from first principles.
Phenomenological hadronic-interaction models are instead developed that require tuning to accelerator data, particularly from the Large Hadron Collider (LHC)~\cite{dEnterria:2011twh,Bierlich:2022pfr,Roesler:2000he,Bopp:2005cr,Pierog:2013ria,Pierog:2025ixr,Ostapchenko:2010vb,Ostapchenko:2024jsg,Ostapchenko:2024myl,Engel:2019dsg}.

The ATLAS Collaboration has performed inclusive measurements with proton-proton ($pp$) data~\cite{STDM-2010-11,STDM-2010-04,STDM-2013-10,STDM-2015-05,STDM-2015-22,STDM-2015-17,STDM-2010-06,STDM-2014-19,STDM-2015-02,STDM-2016-07} and collisions involving lead (Pb) and xenon nuclei~\cite{HION-2011-03,HION-2012-14,HION-2018-17}.
The ALICE~\cite{ALICE:2012fjm,ALICE:2015qqj,ALICE:2017pcy,ALICE:2025woy}, CMS~\cite{CMS-QCD-10-006,CMS-FWD-11-001,CMS-FSQ-13-006,CMS-FSQ-15-001,CMS-FSQ-15-005,CMS-FSQ-16-011,CMS-HIN-16-021,CMS-HIN-17-006,CMS-HIN-23-007,CMS-SMP-23-008}, LHCb~\cite{LHCb:2014wmv,LHCb:2018ehw,LHCb:2021abm,LHCb:2021vww}, and LHCf~\cite{LHCf:2015rcj,LHCf:2017fnw,LHCf:2018gbv,Piparo:2023yam}
collaborations have performed complementary measurements.
Recent LHC results have extended to oxygen-oxygen and neon-neon collisions~\cite{HION-2025-02,ALICE:2025luc,CMS-HIN-25-008,CMS-HIN-25-009}.
However, existing collision systems differ from and poorly constrain hadronic models extrapolated to proton interactions with atmospheric nuclei, which introduces longstanding uncertainties in PeV cosmic-ray results~\cite{Parsons:2019hol,Dembinski:2019uta,Baur:2019cpv,Albrecht:2021yla,Soldin:2021wyv,PierreAuger:2024neu}.
This motivates proton-oxygen ($p$O) collisions that closely resemble high-energy atmospheric interactions~\cite{Brewer:2021kiv,Bruce:2783801,Slupecki:2888741,Waagaard:2025dsi,Leitgeb:2025tep,Maurin:2025gsz,Sharma:2025jhs}.
Inclusive cross-section and particle-production measurements are important to constrain total energy and multiplicity observables in air showers~\cite{Albrecht:2025kbb}.
In addition to collider configuration, LHCb proposes studying these processes at $\sqrt{s_\text{NN}} = 110$ \GeV in their fixed-target system~\cite{BoenteGarcia:2024kba,LHCb:2018ygc,LHCb:2018jry}.
Proton-air cross sections above \TeV center-of-mass energies have only been measured at cosmic-ray observatories~\cite{Baltrusaitis:1984ka,PhysRevLett.70.525,Belov:2006mb,Anatoly:2013lya,EASTOP2009,PierreAuger:2012egl,Nesterova:2017pbw,TelescopeArray:2015oxb,Abbasi:2020chd}.

This Letter presents the first cross-section measurement of charged-particle production in $\sqrt{s_\text{NN}}=9.62$ \TeV $p$O collisions.
This analysis also measures distributions of primary charged particles with $\pt > 500$~\MeV and $|\eta|<2.5$~\cite{atlas-footnote}.
Particles are defined as primary if they have a mean lifetime of $\tau > 300$~ps, either from prompt collision production or decays of promptly produced particles with $\tau < 30$~ps~\cite{STDM-2015-02}.
The LHC delivered a 3.4 \TeV per-nucleon $^{16}\text{O}$ beam traveling in the $+z$ direction and 6.8 \TeV proton beam in the opposite direction.
The dataset recorded by the ATLAS detector in July 2025 corresponds to $634~\mu\text{b}^{-1}$ of integrated luminosity.
The mean number of inelastic interactions per bunch crossing (pileup) is 0.025.

The ATLAS experiment~\cite{PERF-2007-01,GENR-2019-02} comprises an inner detector (ID)~\cite{IDET-2010-01,PERF-2015-07,IDTR-2019-05,IDET-2020-01}, calorimeters, and a muon spectrometer.
The ID features silicon pixels (Pixel) including the innermost B-layer (IBL)~\cite{IDET-2017-10,PIX-2018-001,PIXE-2023-02}, silicon microstrip (SCT)~\cite{SCTD-2019-01}, and transition radiation tracker (TRT)~\cite{IDET-2015-01} subsystems.
Zero-degree calorimeters (ZDC)~\cite{Avoni:2025lmu} are situated at $z=\pm 140$~m from the interaction point.
A software suite~\cite{SOFT-2022-02} is used for simulation, reconstruction and analysis of events, detector operations, and data acquisition.

Monte Carlo (MC) simulated $p$O events are generated with HIJING 1.38~\cite{Wang:1991hta}.
An alternative sample is generated using the default Angantyr~\cite{Bierlich:2016smv,Bierlich:2018xfw} model in \textsc{Pythia} 8.308~\cite{Bierlich:2022pfr}, where \textsc{EvtGen}~\cite{Lange:2001uf} handles heavy-flavor decays.
These samples undergo detector simulation~\cite{SOFT-2010-01} based on \textsc{Geant4}~\cite{Agostinelli:2002hh}.
Samples of models commonly used in cosmic-ray physics are generated with CRMC~2.2.1~\cite{ulrich_2021_5243165}: DPMJET III 2019-1~\cite{Roesler:2000he,Bopp:2005cr}, EPOS LHC-R~\cite{Pierog:2013ria,Pierog:2025ixr}, QGSJET II-04~\cite{Ostapchenko:2010vb}, QGSJET III~\cite{Ostapchenko:2024jsg,Ostapchenko:2024myl}, and \textsc{Sibyll} 2.3e~\cite{Engel:2019dsg}.

A two-level trigger system~\cite{TRIG-2016-01,TRIG-2022-01} selects events with at least one TRT azimuthal sector above threshold via the Level 1 (L1) ``Fast-OR'' algorithm~\cite{Fratina:1229213}, and at least one reconstructed track in the High-Level Trigger~\cite{TRIG-2019-03}.
Standard data-quality criteria are applied~\cite{DAPR-2018-01}.
Selected events must have one primary vertex (PV) with two or more $\pt > 100$ \MeV tracks~\cite{ATL-PHYS-PUB-2019-015}.
To suppress pileup, events with a second vertex containing five or more associated tracks are vetoed.
Events are not removed if a second vertex has fewer than five tracks, which are dominated by secondary interactions reconstructed as another PV or one PV split into two vertices, where one has few tracks.
The fraction of events rejected by this veto due to these reconstruction effects is estimated in the simulation to be
0.02\% and deemed negligible.

Events must have at least one reconstructed track~\cite{PERF-2015-08,IDTR-2022-04,IDTR-2021-01} satisfying $\pt > 500$ \MeV and $|\eta| < 2.5$.
Tracks must have the number of Pixel (SCT) hits satisfying $N_\text{Pixel}^\text{hits} \geq 1$ $(N_\text{SCT}^\text{hits} \geq 6)$.
Tracks must have an innermost-pixel-layer hit if expected~\cite{track-hits-expected}; if the innermost hit is not expected, the next-to-innermost hit is required if expected.
Tracks must satisfy $\left|d^\text{BL}_0\right| < 1.5$~mm and  $\left|z^\text{BL}_0\cdot \sin\theta\right| < 1.5$~mm impact parameter requirements~\cite{impact-param-footnote}.
Tracks with $\pt > 10$ \GeV must have a track-fit $\chi^2$ probability satisfying $P(\chi^2_\text{trk-fit}) > 0.01$ to suppress tracks with mismeasured $\pt$ from material interactions.
A total of 246 million events comprising 5.11 billion tracks satisfy the selection.

The dominant background is secondary-particle tracks due to hadronic interactions, photon conversions, and decays of long-lived particles.
This is estimated by fitting simulated track $d_0^\text{BL}$ distributions without the $|d_0^\text{BL}|<1.5$~mm requirement to the secondaries-dominated sideband $4 < |d_0^\text{BL}| < 9.5$~mm~\cite{STDM-2015-02}.
The fit scales the secondaries' normalization by $1.38 \pm 0.23$, yielding a background fraction of $f_\text{sec} = (2.5\pm 0.4)\%$.
Uncertainties in $f_\text{sec}$ are estimated by repeating this procedure using \textsc{Pythia} instead of HIJING simulation.
Additional subdominant uncertainties arise from varying the sideband fit range and reweighting the simulated $d_0^\text{BL}$ distributions to match data in the $|d_0^\text{BL}|<1.5$~mm region.
Charged particles satisfying $30<\tau<300$~ps are considered background, which removes strange baryons that have low reconstruction efficiencies~\cite{STDM-2015-02}.
The fraction of reconstructed tracks due to strange baryons, $f_\text{sb}$, is estimated via HIJING to be negligible on average, but rises to $f_\text{sb} = (3\pm 1)\%$ for $\pt>20$ \GeV, where the difference relative to \textsc{Pythia} is taken as the uncertainty.
Random combinations of silicon hits producing fake-track backgrounds are estimated by using simulation to have a negligible rate $f_\text{fake}<0.01\%$.
Beam-induced backgrounds~\cite{DAPR-2012-01,DAPR-2014-01,DAPR-2021-02} are estimated by replacing the nominal trigger with a ZDC-based trigger that selects events with only one beam filled; this reduces the rate by four orders of magnitude so such backgrounds are deemed negligible.
Cosmic-ray backgrounds are negligible compared with the event rate~\cite{STDM-2015-02}.

The number of events is corrected for the efficiencies of the trigger $\epsilon_\text{trig}$ and vertex $\epsilon_\text{vtx}$ requirements  using weights $w_\text{evt} = 1/(\epsilon_\text{vtx}\cdot \epsilon_\text{trig})$.
These are determined as a function of the number of selected tracks with the $z^\text{BL}_0 \sin\theta$ requirement removed,  $n_\text{sel}^\text{no-z}$.
The trigger efficiency is measured in data relative to a ZDC-based reference trigger, and is found to be 97.8\% (99.4\%) for $n_\text{sel}^\text{no-z} = 1 (2)$, plateauing at 99.8\% for $n_\text{sel}^\text{no-z}\geq 3$, with negligible dependence on track $\pt$ and $\eta$.
Repeating this procedure using a reference trigger based on random L1 hits results in differences of at most 0.3\% and is taken as the uncertainty.
The vertex efficiency is measured from data as the ratio of the number of selected events with a reconstructed vertex to that with the PV requirement removed, whose uncertainties arising from beam-induced backgrounds are negligible~\cite{STDM-2015-02}.
This is found to be 91.7\% for $n_\text{sel}^\text{no-z} = 1$ and rises to 100\% for $n_\text{sel}^\text{no-z} \geq 2$.
For events with $n_\text{sel}^\text{no-z} = 1$, $\epsilon_\text{vtx}$ is also determined differentially in the selected track $\eta$, rising from 81\% at $|\eta| = 2.5$ to 93\% at $|\eta|=0$.
The integrated trigger (vertex) efficiency is $99.6\%$ ($99.5\%$).

Track reconstruction efficiencies are defined via simulation by $\epsilon_\text{trk}(\pt, \eta) = N^\text{match}_\text{reco} / N_\text{gen}$, where $N_\text{gen}$ is the number of generated primary particles in a $(\pt, \eta)$ bin and $N^\text{match}_\text{reco}$ is the number of reconstructed tracks matched~\cite{track-matching-footnote} to a generated primary particle that can be outside kinematic acceptance.
This accounts for resolution effects that cause primary particles to migrate outside the kinematic range.
The passive material between the Pixel and SCT in the $|\eta|>1.5$ region (PP0) is difficult to simulate accurately, so corrections of up to 3\% in $\epsilon_\text{trk}$ are derived by comparing the efficiency to extend a track reconstructed in the Pixel into the SCT for data and simulation~\cite{STDM-2015-02}.
The simulation imperfectly describes the $\eta$ dependence of the mean number of IBL hits $\langle N_\text{IBL}\rangle$, so data-to-MC corrections of up to 2\% derived from $\langle N_\text{IBL}^\text{MC}\rangle / \langle N_\text{IBL}^\text{data}\rangle$ are applied to $\epsilon_\text{trk}$.
The corrected efficiencies range from 86\% for $|\eta|<1$ to 63\% at $|\eta|>2.4$.
Track distributions are corrected as a function of $\pt$ and $\eta$ using weights $w_\text{trk}(\pt, \eta) = [1 - f_\text{sec}(\pt, \eta) - f_\text{sb}(\pt) ]/\epsilon_\text{trk}(\pt, \eta)$.

The following charged-particle distributions are measured: multiplicity $\frac{1}{N_\text{ev}}\frac{\mathrm{d}N_\text{ev} }{ \mathrm{d}n_\text{ch}}$, transverse momentum $\frac{1}{N_\text{ev}}\frac{1}{ 2\pi \pt }\frac{\mathrm{d}^2 N_\text{ch} }{ \mathrm{d}\eta\mathrm{d}\pt}$~\cite{pt-distribution-footnote}, mean $\pt$ ($\langle \pt\rangle $) versus $n_\text{ch}$, and pseudorapidity $\frac{1}{N_\text{ev}}\frac{\mathrm{d}N_\text{ch} }{ \mathrm{d}\eta}$.
Here, $n_\text{ch}$ is the charged-particle multiplicity in an event, $N_\text{ev}$ is the total number of corrected events, and $N_\text{ch}$ is the total number of charged particles in all events.
Bayesian unfolding~\cite{Richardson:1972hli,Lucy:1974yx,DAgostini:1994fjx} is applied to correct the multiplicity of selected tracks, $n_\text{sel}$, to that of charged particles $n_\text{ch}$.
Unfolding also corrects the $\pt$ distribution of tracks to that of charged particles.
For the dependence of $\langle \pt\rangle$ on $n_\text{ch}$, unfolding separately corrects $\sum_i\pt^i$ (sum over all tracks and events) versus $n_\text{sel}$ and the total number of tracks in all events versus $n_\text{sel}$, before taking the ratio.
The $\eta$ distribution is corrected using only weights as kinematic migrations in $\eta$ are negligible.

Systematic uncertainties in track reconstruction efficiencies arise from 5\%, 10\%, and 25\% uncertainties in the amount of passive material for the ID, IBL, and PP0, respectively~\cite{PERF-2015-07}.
Simulated samples with passive material increased by these amounts are used to reevaluate $\epsilon_\text{trk}$, whose variations are added linearly to account for possible correlated missing material in the simulation.
This yields the following uncertainties in the measured distributions: 0.8\% to 3.6\% from $|\eta|<1$ to $|\eta|>2.3$ regions, 1.5\% to 18\% from low $\pt$ to $\pt >30$ \GeV, and 2\% to 18\% from low $n_\text{ch}$ to $n_\text{ch}>150$.
Uncertainties in $\epsilon_\text{trk}$ also arise from differences in the number of detector hits between simulation and data, which are estimated from data-to-MC differences in the ratio
$n_\text{sel} (\pt, \eta) / n_\text{relax}(\pt, \eta)$.
Here, $n_\text{relax}$ is the number of tracks with the $N_\text{Pixel}^\text{hits}, N_\text{SCT}^\text{hits}, P(\chi^2_\text{trk-fit})$ requirements relaxed to $\geq 0$, $\geq5$, and  $\geq0$, respectively.
The differences are negligible on average, but reach 4\% for $30<\pt < 50$ \GeV and rise further above this.
Differences of 0.1\% to 0.3\% when calculating $\langle N_\text{IBL}^\text{MC}\rangle / \langle N_\text{IBL}^\text{data}\rangle$ corrections using \textsc{Pythia} instead of HIJING are added in quadrature to $\epsilon_\text{trk}$ uncertainties.

Further uncertainties arise from the simulated detector response being sensitive to differences in generator particle composition, where HIJING (\textsc{Pythia}) more accurately describes high-$n_\text{ch}$ ($\pt$ and $\eta$) data.
These generator-dependence uncertainties are estimated from the unfolding procedure not reproducing generator distributions when applied to \textsc{Pythia} events, which are found to be:
1\% to 20\% from low to high $n_\text{ch}$,
0.5\% to 23\% from low to high $\pt$,
0.5\% for the $\eta$ distribution,
and up to $2.5\%$ (0.25\%) for $n_\text{ch}\leq 10$ ($n_\text{ch} > 10$) in the $\langle \pt\rangle$ versus $n_\text{ch}$ distribution.
This has a 1.4\% impact on $N_\text{ev}$ and is its dominant source of uncertainty.
Uncertainties also arise from simulated longitudinal PV positions being reweighted to match data, which is estimated by removing the pileup veto, resulting in a 0.3\% impact on average, rising to 1\% for $|\eta|> 2.4$.

\begin{figure}
\centering
\begin{subfigure}{0.49\textwidth}
\includegraphics[width=\textwidth]{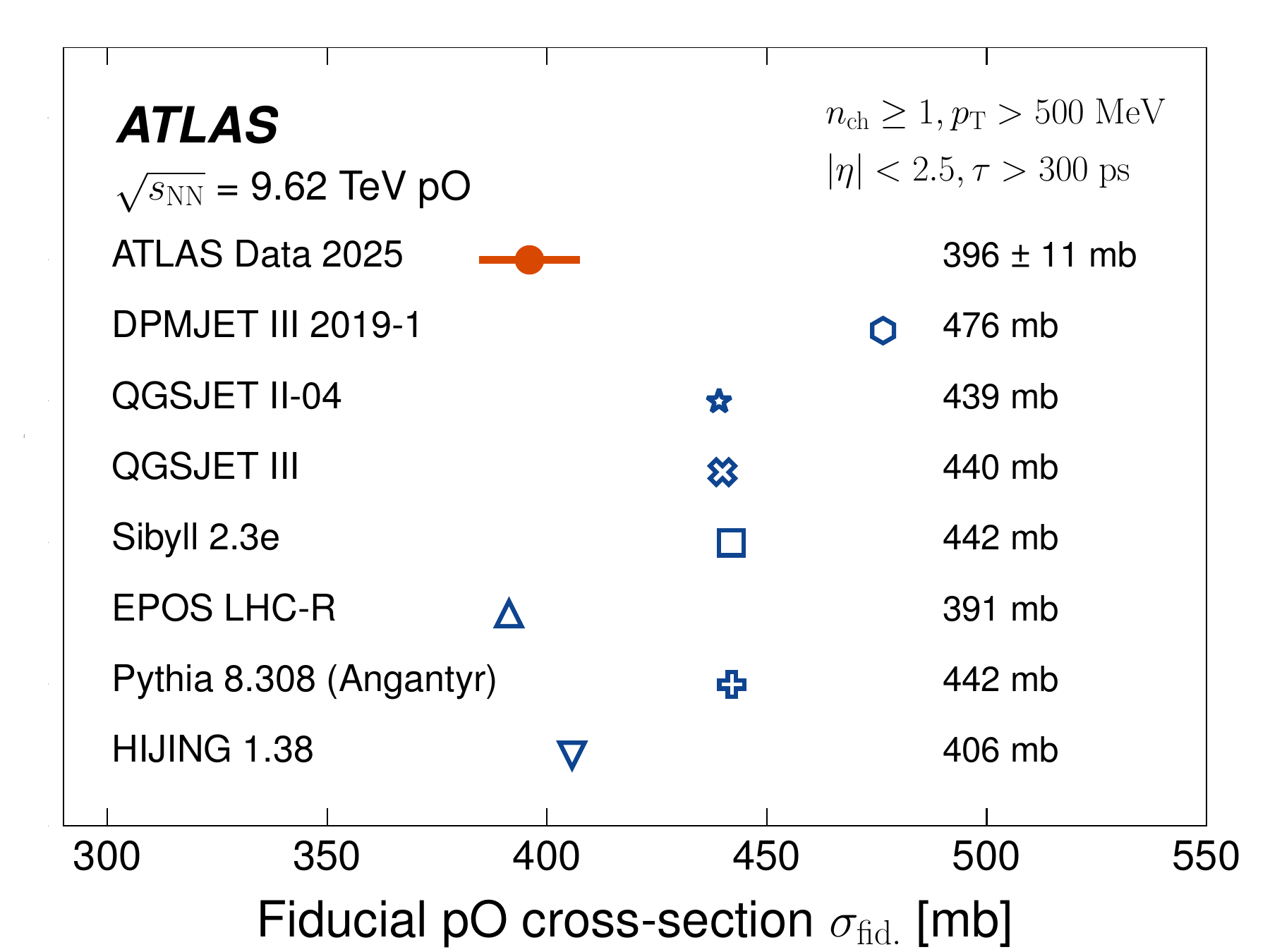}
\caption{\label{pO_xsec_exp_theory_fid}}
\end{subfigure}%
\begin{subfigure}{0.5\textwidth}
\includegraphics[width=\textwidth]{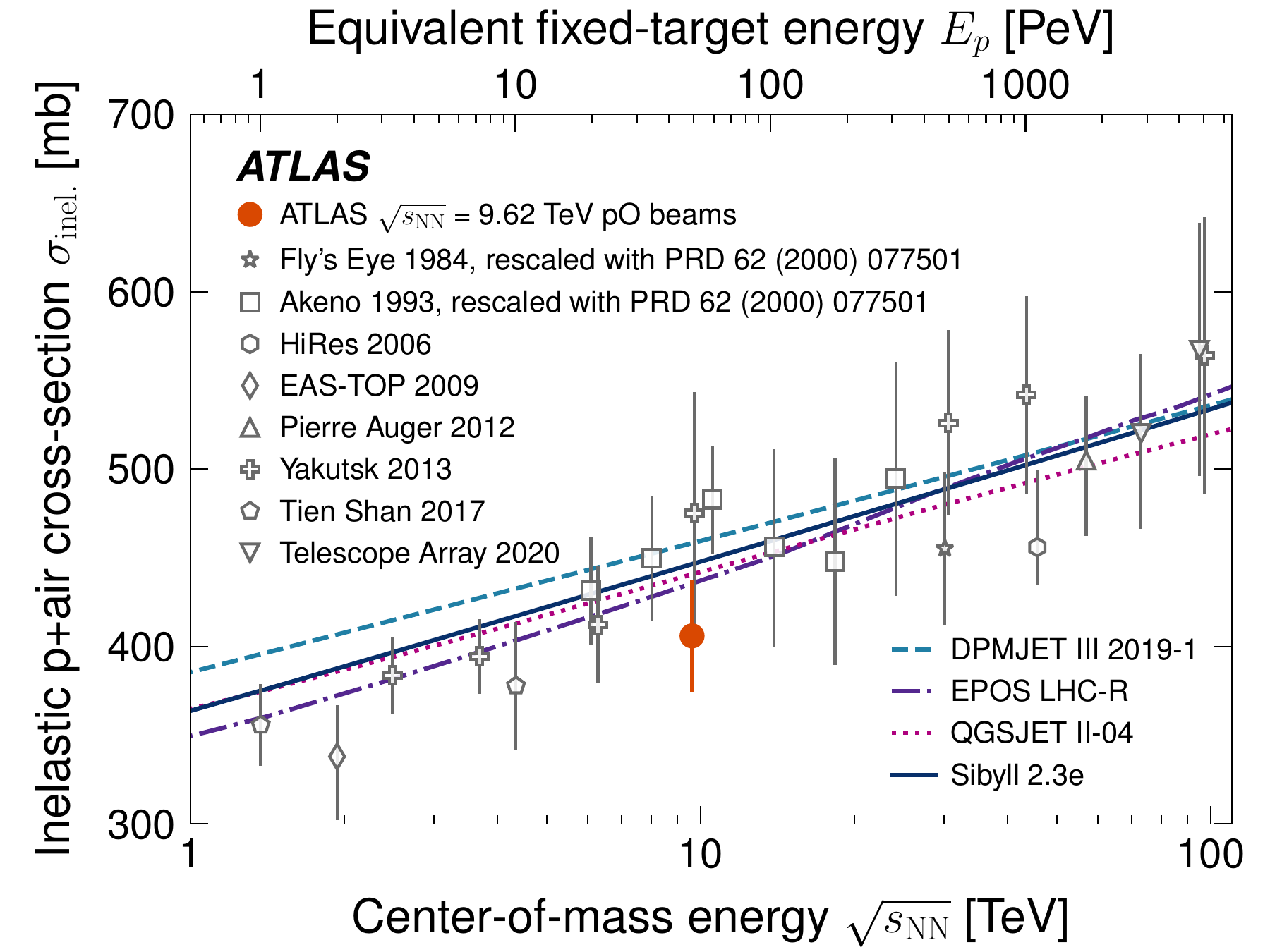}
\caption{\label{fig:pO_xsec_cosmicray-paper}}
\end{subfigure}
\caption{(a) Fiducial $p$O cross-section measurement (filled marker) compared with predictions (open markers) by HIJING 1.38~\cite{Wang:1991hta}, \textsc{Pythia} 8.308~\cite{Bierlich:2022pfr}, DPMJET III 2019-1~\cite{Roesler:2000he}, EPOS LHC-R~\cite{Pierog:2025ixr}, \textsc{Sibyll} 2.3e~\cite{Engel:2019dsg}, QGSJET II-04~\cite{Ostapchenko:2010vb} and III~\cite{Ostapchenko:2024jsg,Ostapchenko:2024myl}. The error bar is the quadrature sum of experimental and luminosity uncertainties. (b) The proton-air cross-section measurement (filled marker) is shown versus center-of-mass $\sqrt{s_\text{NN}}$ and equivalent fixed-target  $E_p$ energy with predictions (lines) by DPMJET III 2019-1, EPOS LHC-R, QGSJET II-04, and \textsc{Sibyll} 2.3e. Observatory results (open markers) are by Fly's Eye~\cite{Baltrusaitis:1984ka}, Akeno~\cite{PhysRevLett.70.525}, HiRes~\cite{Belov:2006mb}, Yakutsk~\cite{Anatoly:2013lya}, EAS-TOP~\cite{EASTOP2009}, Auger~\cite{PierreAuger:2012egl}, Tien Shan~\cite{Nesterova:2017pbw}, Telescope Array~\cite{TelescopeArray:2015oxb,Abbasi:2020chd}. Rescaled values for Refs.~\cite{Baltrusaitis:1984ka,PhysRevLett.70.525} are from Ref.~\cite{Block:2000pg}.}
\end{figure}

\begin{figure}
\centering
\begin{subfigure}{0.47\textwidth}
\includegraphics[width=\textwidth]{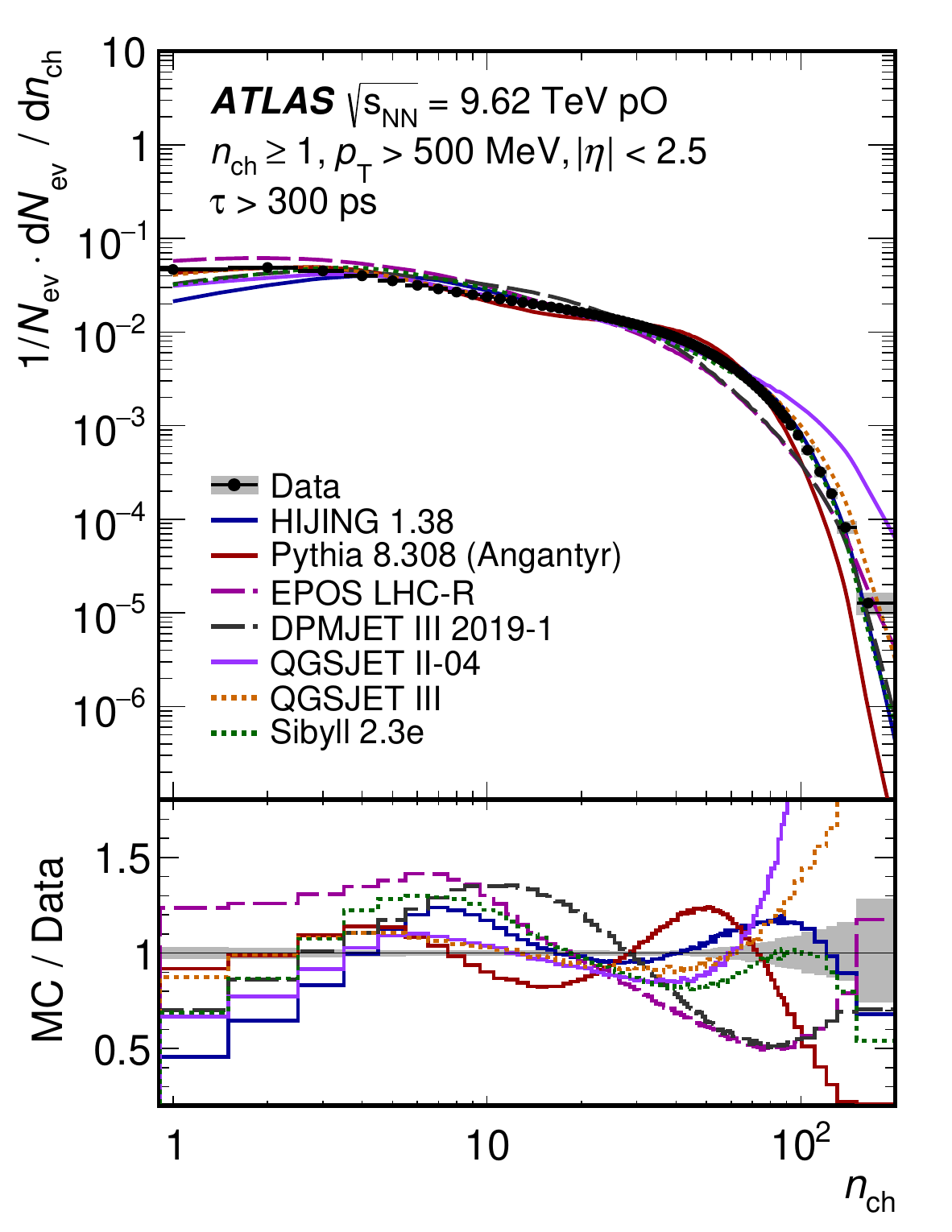}
\caption{\label{fig:unfolded-data-paper-nch}}
\end{subfigure}
\begin{subfigure}{0.47\textwidth}
\includegraphics[width=\textwidth]{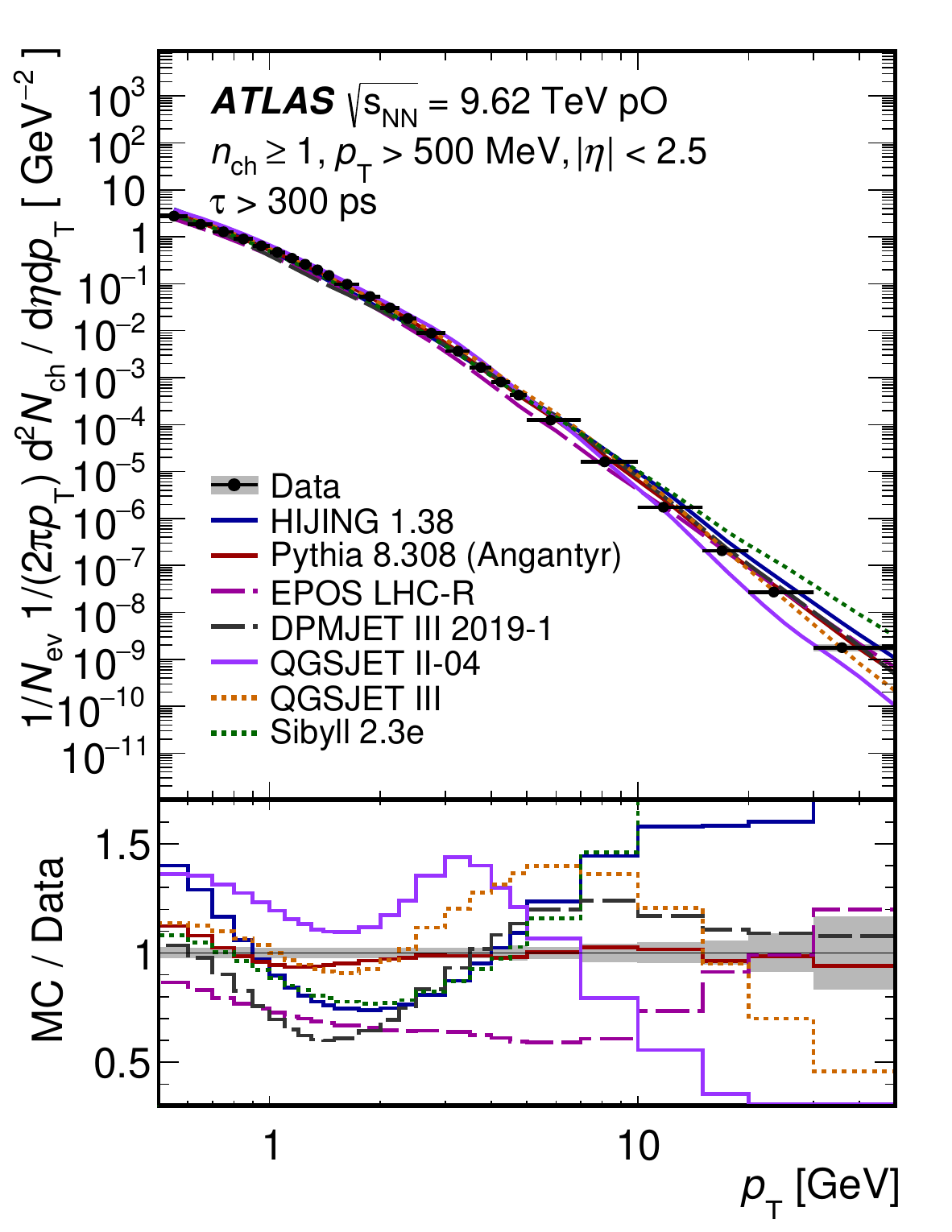}
\caption{\label{fig:unfolded-data-paper-pt}}
\end{subfigure}
\begin{subfigure}{0.47\textwidth}
\includegraphics[width=\textwidth]{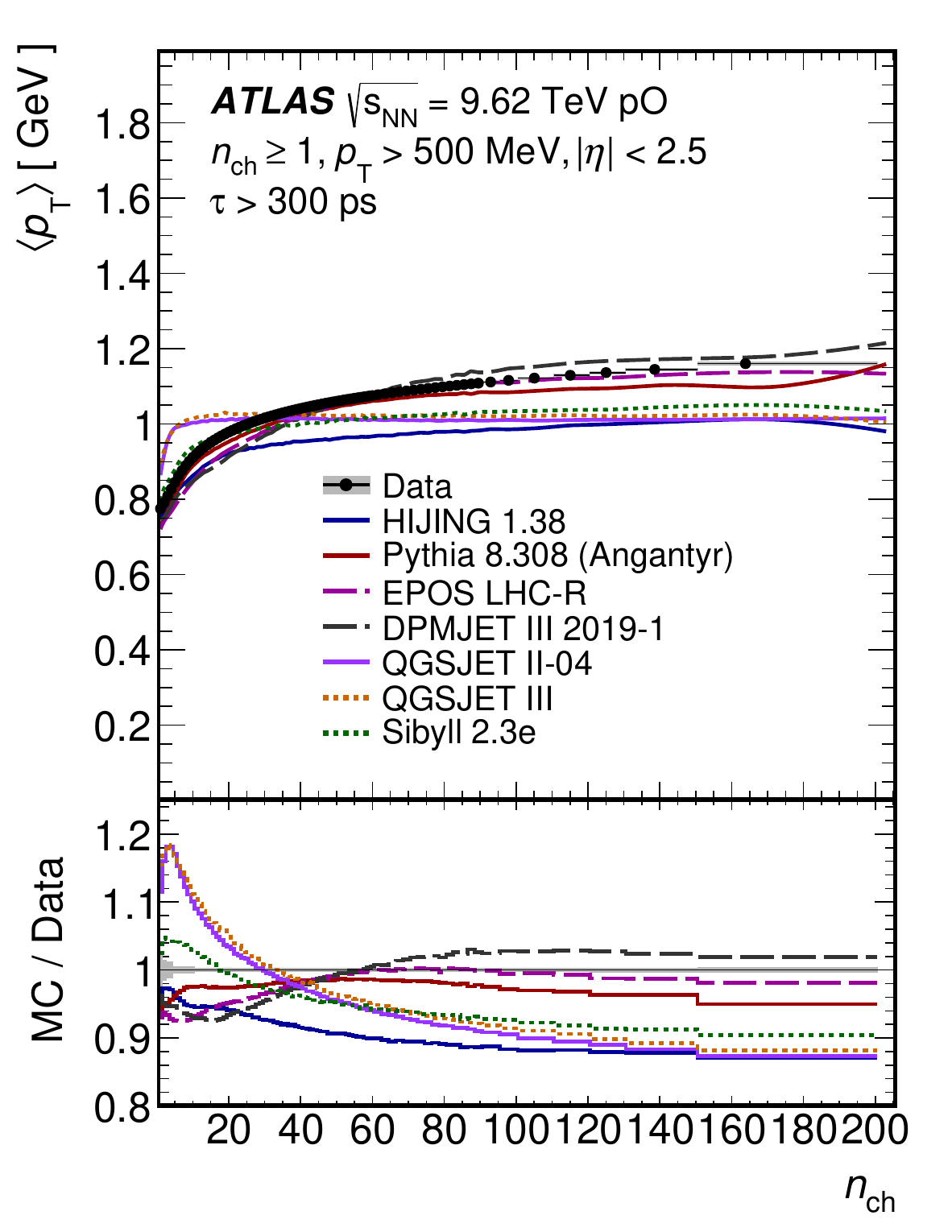}
\caption{\label{fig:unfolded-data-paper-ptnch}}
\end{subfigure}
\begin{subfigure}{0.47\textwidth}
\includegraphics[width=\textwidth]{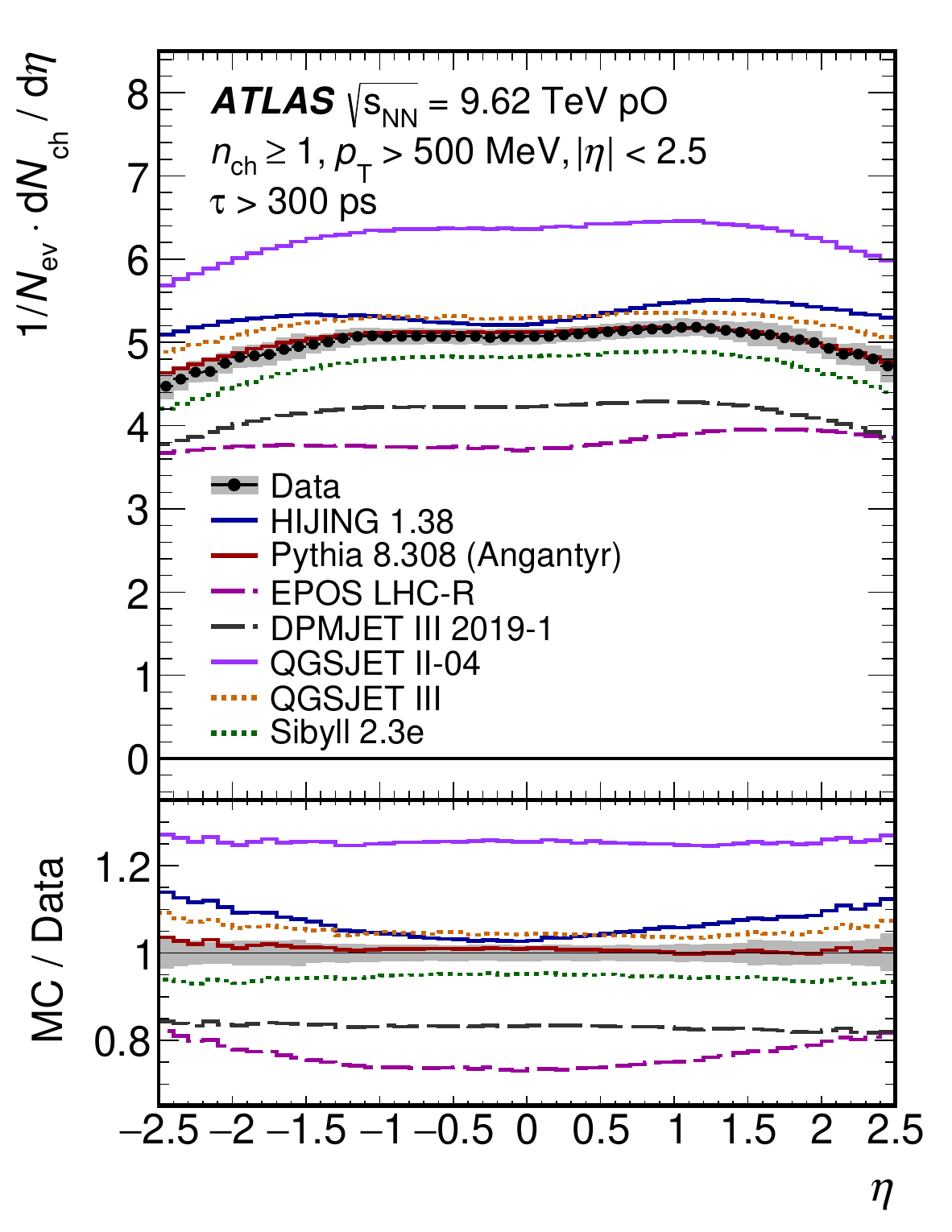}
\caption{\label{fig:unfolded-data-paper-eta}}
\end{subfigure}
\caption{\label{fig:unfolded-data-paper}Primary charged-particle distributions in the fiducial region for data (black points), where the shaded band denotes the combined statistical and systematic uncertainties. The $x$-value in each bin corresponds to the bin centroid. Overlaid are model predictions by HIJING 1.38 (blue), \textsc{Pythia} 8.308 Angantyr (red), EPOS LHC-R (pink), DPMJET III 2019-1 (black), QGSJET II-04 (purple), QGSJET III (orange), and \textsc{Sibyll} 2.3e (green). The lower panel shows the ratio of MC predictions to the data. }
\end{figure}

The fiducial cross section, $\sigma_\text{fid.} = N_\text{ev}/\mathcal{L}$, is measured in a region with at least one primary charged particle satisfying $\pt > 500$ \MeV and $|\eta|<2.5$.
The total number of corrected events is $N_\text{ev} = (251\pm 4)\times 10^6$.
The luminosity and its 2.3\% uncertainty, $\mathcal{L} = 634 \pm 15~\mu$b$^{-1}$, is derived using procedures detailed in Ref.~\cite{DAPR-2021-01} with the LUCID-2 detector~\cite{LUCID2}.
This results in a measured fiducial cross section of
\begin{equation}
\sigma_\text{fid.}^{p\text{O}} =  396 \pm 6~(\text{exp.}) \pm 9~(\text{lumi.})~\text{mb}.
\end{equation}
Statistical uncertainties are negligible and the luminosity calibration is the dominant uncertainty.
Model predictions in the fiducial region are evaluated using \textsc{Rivet}~\cite{Bierlich:2019rhm}; strange baryons $(\Sigma^\pm, \Xi^-, \Omega^-)$ are explicitly removed, which has at most a 0.15\% impact.
Figure~\ref{pO_xsec_exp_theory_fid} shows this measurement is compatible with EPOS and HIJING, but is lower by at least three standard deviations than other models.
Models can overpredict hadronic cross sections due to underestimated diffractive events escaping detector acceptance~\cite{STDM-2015-05,CMS-FSQ-15-005}, where EPOS compatibility with data possibly suggests better diffractive modeling.

The cross section is extrapolated to the full inelastic phase space, $\sigma_\text{inel.} = \sigma_\text{fid.}/A$, using the mean fiducial acceptance $A$ determined from DPMJET III 2019-1 (95.9\%), EPOS LHC-R (84.2\%), \textsc{Pythia} 8.308 (89.1\%), QGSJET II-04 (89.9\%), and \textsc{Sibyll} 2.3e (92.5\%).
This yields $A = (90 \pm 6)\%$, where the maximum difference from the mean is taken as the theory uncertainty.
The measured cross section is
\begin{align}
\sigma_\text{inel.}^{p\text{O}} = 438 \pm 6~(\text{exp.}) \pm 10~(\text{lumi.}) \pm 30~(\text{th.})~\text{mb}.
\label{eq:pO-inel-meas}
\end{align}
Hadronic modeling is the dominant uncertainty arising from the model-dependent extrapolation outside kinematic acceptance.
This result is compatible with EPOS (465~mb) and HIJING (413~mb), while DPMJET III 2019-1 (497~mb), \textsc{Pythia} (496~mb), QGSJET II-04 (489~mb), QGSJET III (487~mb), and \textsc{Sibyll} (478~mb) overpredict data by up to two standard deviations.

The inelastic proton-air cross section is inferred from Eq.~\eqref{eq:pO-inel-meas} using an MC-based scale factor, $\sigma_\text{inel.}^{p+\text{air}} = f_\text{air}\cdot \sigma_\text{inel.}^{p\text{O}}$, where $f_\text{air}= 0.78 \cdot (\sigma_\text{MC}^{p\text{N}} / \sigma_\text{MC}^{p\text{O}}) + 0.22$ assumes air comprises 78\% nitrogen and 22\% oxygen~\cite{argon-footnote}.
The ratio of the proton-nitrogen ($p$N) to $p$O cross sections, $\sigma_\text{MC}^{p\text{N}} / \sigma_\text{MC}^{p\text{O}} = 0.90 \pm 0.02$, is taken as the mean of the values from DPMJET III 2019-1 (0.899), EPOS LHC-R (0.918), \textsc{Pythia} 8.308 (0.921), QGSJET II-04 (0.888), and \textsc{Sibyll} 2.3e (0.895).
The 1.8\% maximum difference from the mean is added in quadrature to the theory uncertainty.
This yields $f_\text{air} = 0.93 \pm 0.02$ and a measured proton-air cross section of
\begin{equation}
\sigma^{p+\text{air}}_\text{inel.} = 406 \pm 6~(\text{exp.}) \pm 9~(\text{lumi.}) \pm 28~(\text{th.})~\text{mb}.
\end{equation}
Figure~\ref{fig:pO_xsec_cosmicray-paper} compares this result with predictions and noncollider measurements at different energies.
LHC collisions at $\sqrt{s_\text{NN}}=9.62$ \TeV correspond to a cosmic-ray proton energy of 49~PeV in the target-nucleus rest frame.
This result is compatible with measurements at similar energies by Akeno~\cite{PhysRevLett.70.525} rescaled with Ref.~\cite{Block:2000pg} and Yakutsk~\cite{Anatoly:2013lya}, and within two standard deviations of models.

Figure~\ref{fig:unfolded-data-paper} shows charged-particle distributions in the fiducial region compared with model predictions~\cite{hepdata-169575}.
The dominant uncertainties in data arise from unfolding generator dependence and track reconstruction.

Figure~\ref{fig:unfolded-data-paper-nch} shows the $n_\text{ch}$ distribution, which probes the interplay between soft and hard particle production.
No model correctly describes $n_\text{ch}$, so the precise data with 1.5\% to 28\% uncertainties are key to resolving these discrepancies.
Hadronization tuning is important for $n_\text{ch}$ modeling~\cite{Albrecht:2021yla}, while multiple parton interactions (MPI)~\cite{Sjostrand:2004pf} are significant at high $n_\text{ch}$.
Angantyr (QGSJET III) are the least discrepant with data at low $n_\text{ch} < 80$, but underestimate (overestimate) data by an order of magnitude at high $n_\text{ch}$, motivating retuned MPI or color-coherence modeling~\cite{Lonnblad:2023stc}.
EPOS treats energy conservation in particle production and cross sections consistently~\cite{Pierog:2025ixr}, which suppresses MPI and narrows $n_\text{ch}$ distributions to better describe $pp$ data~\cite{Pierog:2017awp}, but the extrapolation to $p$O collisions predicts too many low-$n_\text{ch}$ events.

Figure~\ref{fig:unfolded-data-paper-pt} shows the \pt distribution, which probes the transition from soft particle production and hadronization at low \pt to perturbative hard processes at high \pt.
Data uncertainties range from 2.5\% to 17\%, which constrains the considerable spread in model predictions.
Angantyr is the least discrepant with data, being within 12\% throughout.
DPMJET, HIJING and \textsc{Sibyll} predict harder spectra while QGSJET predicts softer spectra than data.

Figure~\ref{fig:unfolded-data-paper-ptnch} shows the dependence of $\langle\pt\rangle$ on $n_\text{ch}$.
The increasing trend in data indicates color-coherence effects in dense parton environments, which reduces extra particles produced with every additional MPI while an equally large \pt is shared among the hadrons~\cite{Lonnblad:2023stc}.
These collective effects are modeled using string fusion in DPMJET, core-corona method in EPOS, and color reconnection in \textsc{Pythia}, which describe the rising trend in data within 8\%.
QGSJET and \textsc{Sibyll} do not model color-coherence effects so predict little dependence on $n_\text{ch}$, in tension with data.

Figure~\ref{fig:unfolded-data-paper-eta} shows the $\eta$ distribution.
The 1.9\% to 4.4\% uncertainties in data constrain the peak particle density of $5.2\pm 0.1$ at $\eta=1.1$, falling to $4.7 \pm 0.2$ ($4.5\pm 0.2$) at $\eta>2.4 (<-2.4)$.
The balance of soft and hard MPI controls the distribution width and height while the $p$O system impacts the forward-backward asymmetry.
Angantyr describes the data within uncertainties.
DPMJET and \textsc{Sibyll} model soft and hard processes separately, which describe the shape in $p$O data despite being too narrow in $pp$ collisions~\cite{Pierog:2017awp}.
The semi-hard approach~\cite{Pierog:2025ixr} in EPOS produces analogs of initial and final-state radiation~\cite{Albrecht:2025kbb}, which
broadens the $\eta$ shape to improve $pp$ modeling~\cite{Pierog:2017awp}, but is too broad when extrapolated to $p$O collisions.

\FloatBarrier
In summary, ATLAS has measured charged-particle production in $\sqrt{s_\text{NN}}=9.62$ \TeV proton-oxygen collisions.
The fiducial cross section, $\sigma_\text{fid.}^{p\text{O}} = 396 \pm 6~(\text{exp.}) \pm 9~(\text{lumi.})~\text{mb}$, and inelastic proton-air cross section, $\sigma^{p+\text{air}}_\text{inel.} = 406 \pm 6~(\text{exp.}) \pm 9~(\text{lumi.}) \pm 28~(\text{th.})~\text{mb}$, favor lower values among predictions.
Models poorly describe the particle multiplicity while kinematic spectra prefer the Angantyr model.
No model can consistently describe all reported results, so these precise data provide valuable input for generator tuning with significant potential to improve the modeling of cosmic-ray interactions in Earth's atmosphere.

\section*{Acknowledgments}
%

%

%
%

%
%

We thank CERN for the very successful operation of the LHC and its injectors, as well as the support staff at
CERN and at our institutions worldwide without whom ATLAS could not be operated efficiently.

The crucial computing support from all WLCG partners is acknowledged gratefully, in particular from CERN, the ATLAS Tier-1 facilities at TRIUMF/SFU (Canada), NDGF (Denmark, Norway, Sweden), CC-IN2P3 (France), KIT/GridKA (Germany), INFN-CNAF (Italy), NL-T1 (Netherlands), PIC (Spain), RAL (UK) and BNL (USA), the Tier-2 facilities worldwide and large non-WLCG resource providers. Major contributors of computing resources are listed in Ref.~\cite{ATL-SOFT-PUB-2026-001}.

We gratefully acknowledge the support of ANPCyT, Argentina; YerPhI, Armenia; ARC, Australia; BMWFW and FWF, Austria; ANAS, Azerbaijan; CNPq and FAPESP, Brazil; NSERC, NRC and CFI, Canada; CERN; ANID, Chile; CAS, MOST, NSFC and FRFCU, China; Minciencias, Colombia; MEYS CR, Czech Republic; DNRF and DNSRC, Denmark; IN2P3-CNRS and CEA-DRF/IRFU, France; SRNSFG, Georgia; BMFTR, HGF and MPG, Germany; GSRI, Greece; RGC and Hong Kong SAR, China; ICHEP and Academy of Sciences and Humanities, Israel; INFN, Italy; MEXT and JSPS, Japan; CNRST, Morocco; NWO, Netherlands; RCN, Norway; MNiSW, Poland; FCT, Portugal; MNE/IFA, Romania; MSTDI, Serbia; MSSR, Slovakia; ARIS and MVZI, Slovenia; DSI/NRF, South Africa; MICIU/AEI, Spain; SRC and Wallenberg Foundation, Sweden; SERI, SNSF and Cantons of Bern and Geneva, Switzerland; NSTC, Taipei; TENMAK, T\"urkiye; STFC/UKRI, United Kingdom; DOE and NSF, United States of America.

Individual groups and members have received support from BCKDF, CANARIE, CRC and DRAC, Canada; CERN-CZ, FORTE and PRIMUS, Czech Republic; COST, ERC, ERDF, Horizon 2020 and Marie Sk{\l}odowska-Curie Actions, European Union; Investissements d'Avenir Labex, Investissements d'Avenir Idex and ANR, France; DFG and AvH Foundation, Germany; Herakleitos, Thales and Aristeia programmes co-financed by EU-ESF and the Greek NSRF, Greece; BSF-NSF and MINERVA, Israel; NCN and NAWA, Poland; La Caixa Banking Foundation, CERCA and AGAUR programs from Generalitat de Catalunya and PROMETEO and GenT Programmes Generalitat Valenciana, Spain; G\"{o}ran Gustafssons Stiftelse, Sweden; The Royal Society and Leverhulme Trust, United Kingdom; Eric and Wendy Schmidt Fund for Strategic Innovation, United States of America.

In addition, individual members wish to acknowledge support from Chile: Agencia Nacional de Investigaci\'on y Desarrollo (ANID FONDECYT reg. 1230987, FONDECYT 1230812, FONDECYT 1240864, Fondecyt 3240661, Fondecyt Regular 1240721); China: Fundamental Research Funds for the Central Universities (010-63263105), Chinese Ministry of Science and Technology (MOST-2023YFA1605700, MOST-2023YFA1609300), National Natural Science Foundation of China (NSFC 12275265, NSFC-W2543005); Czech Republic: Czech Science Foundation (GACR - 24-11373S), Ministry of Education Youth and Sports (ERC-CZ-LL2327, FORTE CZ.02.01.01/00/22\_008/0004632), PRIMUS Research Programme (PRIMUS/21/SCI/017); EU: H2020 European Research Council (ERC - 101002463); European Union: European Research Council (BARD No. 101116429, ERC - 101219398, ERC - 948254, ERC 101089007), European Regional Development Fund (HE COFUND GA No.101081355, ERDF), Marie Sklodowska-Curie Actions (GAP-101168829); France: Agence Nationale de la Recherche (ANR-21-CE31-0013, ANR-22-EDIR-0002, ANR-24-CE31-0504-01); Germany: Deutsche Forschungsgemeinschaft (DFG - 469666862); China: Research Grants Council (GRF); Italy: Istituto Nazionale di Fisica Nucleare (LHC-MIUR - 28003/2025), Ministero dell'Università e della Ricerca (NextGenEU 153D23001490006 M4C2.1.1, NextGenEU I53D23000820006 M4C2.1.1, SOE2024\_0000023); Japan: Japan Society for the Promotion of Science (JSPS KAKENHI  JP25H0063, JSPS KAKENHI JP22H01227, JSPS KAKENHI JP22H04944, JSPS KAKENHI JP22KK0227, JSPS KAKENHI JP24K23939, JSPS KAKENHI JP24KK0251, JSPS KAKENHI JP25H00650, JSPS KAKENHI JP25H01291, JSPS KAKENHI JP25K01011, JSPS KAKENHI JP25K01023); Poland: Polish National Science Centre (NCN 2021/42/E/ST2/00350, NCN OPUS 2023/51/B/ST2/02507, NCN OPUS nr 2022/47/B/ST2/03059, NCN UMO-2019/34/E/ST2/00393, UMO-2022/47/O/ST2/00148, UMO-2023/49/B/ST2/04085, UMO-2023/51/B/ST2/00920, UMO-2024/53/N/ST2/00869); Spain: Agència de Gestió d'Ajuts Universitaris i de Recerca. (AGAUR - 2023 BP 00141), Ministry of Science and Innovation (RYC2019-028510-I, RYC2020-030254-I, RYC2021-031273-I, RYC2022-038164-I), Ministerio de Ciencia, Innovación y Universidades/Agencia Estatal de Investigaci\'on (EU NextGenerationEU (PRTR-C17.I1), PID2022-142604OB-C22); Sweden: Carl Trygger Foundation (Carl Trygger Foundation CTS 22:2312), Swedish Research Council (Swedish Research Council 2023-04654, VR 2021-03651, VR 2022-03845, VR 2022-04683, VR 2023-03403, VR 2024-05451, VR 2025-05940), Knut and Alice Wallenberg Foundation (KAW 2023.0366); Switzerland: Swiss National Science Foundation (SNSF - PCEFP2\_194658); United Kingdom: The Binks Trust, Royal Society (NIF-R1-231091); United States of America: U.S. Department of Energy (ECA DE-AC02-76SF00515), John Templeton Foundation (John Templeton Foundation 63206), Neubauer Family Foundation.

%
%


%
%
%
%
%
%

%
%

%
%
\printbibliography
\clearpage
 
\begin{flushleft}
\hypersetup{urlcolor=black}
{\Large The ATLAS Collaboration}

\bigskip

\AtlasOrcid[0000-0002-6665-4934]{G.~Aad}$^\textrm{\scriptsize 102}$,
\AtlasOrcid[0000-0001-7616-1554]{E.~Aakvaag}$^\textrm{\scriptsize 17}$,
\AtlasOrcid[0000-0002-5888-2734]{B.~Abbott}$^\textrm{\scriptsize 121}$,
\AtlasOrcid[0000-0002-0287-5869]{S.~Abdelhameed}$^\textrm{\scriptsize 83b}$,
\AtlasOrcid[0000-0002-1002-1652]{K.~Abeling}$^\textrm{\scriptsize 54}$,
\AtlasOrcid[0000-0001-5763-2760]{N.J.~Abicht}$^\textrm{\scriptsize 48}$,
\AtlasOrcid[0000-0002-8496-9294]{S.H.~Abidi}$^\textrm{\scriptsize 30}$,
\AtlasOrcid[0009-0003-6578-220X]{M.~Aboelela}$^\textrm{\scriptsize 44}$,
\AtlasOrcid[0000-0002-9987-2292]{A.~Aboulhorma}$^\textrm{\scriptsize 36e}$,
\AtlasOrcid[0000-0001-5329-6640]{H.~Abramowicz}$^\textrm{\scriptsize 154}$
% ...
\bigskip
\\

% ...
$^{17}$Department for Physics and Technology, University of Bergen, Bergen; Norway.\\
% ...
$^{30}$Physics Department, Brookhaven National Laboratory, Upton NY; United States of America.\\
% ...
$^{36}$$^{(a)}$Facult\'e des Sciences Ain Chock, Universit\'e Hassan II de Casablanca;$^{(b)}$Facult\'{e} des Sciences, Universit\'{e} Ibn-Tofail, K\'{e}nitra;$^{(c)}$Facult\'e des Sciences Semlalia, Universit\'e Cadi Ayyad, LPHEA-Marrakech;$^{(d)}$LPMR, Facult\'e des Sciences, Universit\'e Mohamed Premier, Oujda;$^{(e)}$Facult\'e des sciences, Universit\'e Mohammed V, Rabat;$^{(f)}$Institute of Applied Physics, Mohammed VI Polytechnic University, Ben Guerir; Morocco.\\
% ...
$^{44}$Physics Department, Southern Methodist University, Dallas TX; United States of America.\\
% ...
$^{48}$Fakult\"{a}t Physik, Technische Universit{\"a}t Dortmund, Dortmund; Germany.\\
% ...
$^{54}$II. Physikalisches Institut, Georg-August-Universit\"{a}t G\"ottingen, G\"ottingen; Germany.\\
% ...
$^{83}$$^{(a)}$Khalifa University of Science and Technology, Abu Dhabi;$^{(b)}$New York University Abu Dhabi, Abu Dhabi;$^{(c)}$United Arab Emirates University, Al Ain;$^{(d)}$University of Sharjah, Sharjah; United Arab Emirates.\\
% ...
$^{102}$CPPM, Aix-Marseille Universit\'e, CNRS/IN2P3, Marseille; France.\\
% ...
$^{121}$Homer L. Dodge Department of Physics and Astronomy, University of Oklahoma, Norman OK; United States of America.\\
% ...
$^{154}$Raymond and Beverly Sackler School of Physics and Astronomy, Tel Aviv University, Tel Aviv; Israel.\\
% ...
$^{a}$ Also at Affiliated with an institute formerly covered by a cooperation agreement with CERN.\\
$^{b}$ Also at An-Najah National University, Nablus; Palestine.\\
$^{c}$ Also at Borough of Manhattan Community College, City University of New York, New York NY; United States of America.\\
$^{d}$ Also at Center for Interdisciplinary Research and Innovation (CIRI-AUTH), Thessaloniki; Greece.\\
$^{e}$ Also at Centre of Physics of the Universities of Minho and Porto (CF-UM-UP); Portugal.\\
$^{f}$ Also at CERN, Geneva; Switzerland.\\
% ...

\end{flushleft}


%

%
%

%

%
%
%
%

\end{document}